# Against negative splitting: the case for alternative pacing strategies for elite marathon athletes in official events

G.D. Fernandes*, Victor Maldonado

(*) corresponding author, gdufflis@ttu.edu

Mechanical Engineering department, Texas Tech University

**Objectives**

Negative splitting (i.e., finishing the race faster) is a tactic commonly employed by elite marathon athletes, even though research supporting the strategy is scarce. The presence of pacers allows the main runner to run behind a formation, preserving energy. Our aim is to show that, in the presence of pacers, the most efficient pacing strategy is positive splitting.

**Methods**

We evaluated the performance of an elite marathon runner from an energetic standpoint, including drag values obtained through Computational Fluid Dynamics (CFD). In varying simulations for different pacing strategies, the energy for both the main runner and his pacer were conserved and the total race time was calculated.

**Results**

In order to achieve minimum race time, the main runner must start the race faster and run behind the pacers, and when the pacers drop out, finish the race slower. Optimal race times are obtained when the protected phase is run 2.4 to 2.6% faster than the unprotected phase.

**Conclusion**

Our results provide strong evidence that positive splitting is indeed the best pacing strategy when at least one pacer is present, causing significant time savings in official marathon events.

**Summary Box**

**What is already known on this topic**

Most marathon athletes tend to adopt a negative splitting strategy during races, although this practice is not supported by significant research.

**What this study adds**

If the main runner has pacers available, the optimal drafting/pacing strategy features a positive splitting strategy, contrary to popular belief.

**How this study might affect research, practice or policy**

Coaches and elite marathon runners need to be aware of this discovery, which may lead to reduced race times.

## 1. Introduction

The pacing strategy adopted by elite athletes is a crucial element of performance. In track events, different pacing strategies are usually adopted depending on the distance. For shorter events, such as the 100-m sprint, an "all-out" pacing strategy is common; for long distance events, such as marathons or ultra marathons, a choice of negative, even or positive pacing is possible[1], with elite athletes usually adopting a negative pacing strategy, i.e., increasing the speed during the race. Research has shown that, before 1988, record-setting marathons were run with a positive split, with a sharp decrease in pace at the 25-km mark [2]. Ever since, the fastest runners have used more consistent pacing across all segments of the race, such as Paula Radcliffe in her 2003 world record [3]. One such example is Kenyan runner Eliud Kipchoge, who broke the marathon record in the 2018 Berlin Marathon, and once again in the 2022 edition. Fig. 1-a shows his pacing splits for the 2018 edition, consisting of mostly even but fluctuating pacing, save for the final 2.2 km of the race, for a 2:01:39 finish. In the 2022 edition (Fig. 1-b), a positive splitting strategy is clearly visible, with all 5-km segments after km 20 being run slower than the global average pace (save, again, for the final 2.2 km). Whether this pacing strategy contributed to the 2:01:09 world record remains a question.

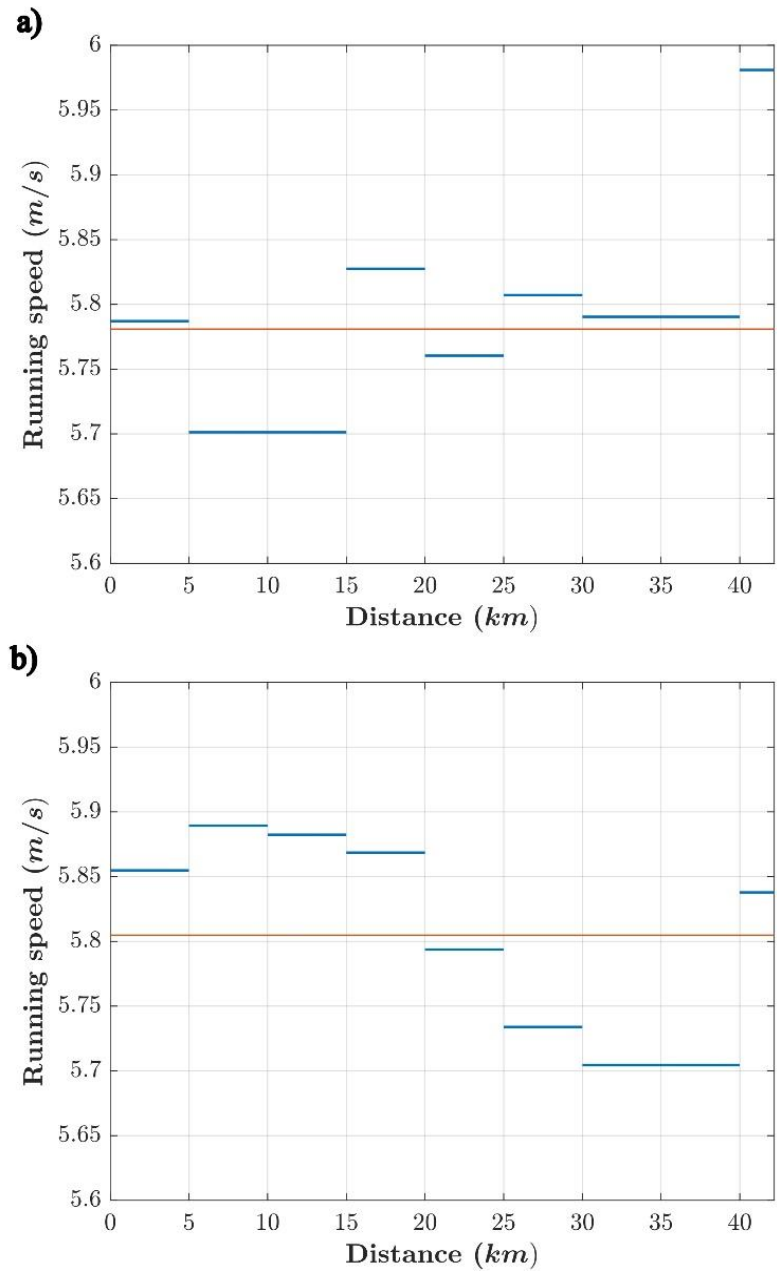

Fig. 1 Eliud Kipchoge's pacing splits for the 2018 (a) and 2022 (b) editions of the Berlin Marathon [4]. The average velocity for each 5-km segment of the race is represented by the blue lines, and the global average velocity is represented by the red line.

Positive splitting has been shown to be detrimental to marathon running performance by causing an increased oxygen consumption, a higher level of fatigue, and an increased rate of perceived exertion[3]. That means fatigue happens faster than when compared to an even-splitting strategy. However, that might not be the case for slight positive pacing, and Kipchoge's performance supports this hypothesis. Still, the most common pacing strategy by elite marathon runners is negative splitting. It is thought to improve prolonged exercise performance by reducing the rate of carbohydrate depletion, lowering excessive oxygen consumption (VO2), and/ or limiting the accumulation of fatigue-related metabolites [2]. However, there is limited evidence that negative splitting actually results in lower race times [5], and pacing strategies are often determined by experience. Curiously, one important aspect of marathon running that directly affects the pacing strategy has never been investigated: the presence of pacers along with the main runner. Usually, marathon record holders are elite athletes who run with two or three runners tasked with assisting the main runner maintain his or her pace. When the pacers can't keep up with the target pace, they leave the race, and the main runner finishes alone. However, the presence of pacers has a secondary benefit that greatly affects performance and is seldom explored to its full potential: drag reduction.

The process of running behind other runners in formation to benefit from lower air resistance is known as drafting and is common in many sports [6–17]. In the context of marathon running, it has been shown that different formations can result in drag reductions in excess of 70% for the main runner [18,19]. The reduced drag results in lower running power, which can be used to either preserve energy during the race or to run at a faster speed with the same energy expenditure. In the 2018 and 2022 editions of the Berlin Marathon, Eliud Kipchoge benefitted from drafting by running in different formations, which contributed to his great performances; however, research has shown that he could have adopted different formations and run at different speeds to finish the races even faster [20]. In particular, it was shown that one specific drafting/pacing strategy could even lead him to a historic sub 2-hour finish. In October 2023, when fellow Kenyan runner Kelvin Kiptum broke the record by running the Chicago Marathon in 2:00:35 (without benefitting from drafting at all), there was growing speculation that the breaking of the 2-hour barrier is a matter of time. In fact, research has shown that proper use of drafting could lead Kiptum to break the mark by a wide margin [21].

**Considerations about negative splitting**

The drafting strategy must always be evaluated along with the pacing strategy. The runner (s) in front of the formation are subject to higher drag, and therefore spend more energy than the protected runner. After the pacers leave, the main runner is then subject to significant drag. In order to optimize the performance, it is ideal to keep at least one pacer in the race for as long as possible (so the main runner can benefit from reduced drag for longer). As it is shown in section 2, the running power roughly increases with the square of the running speed. In other words, running slightly faster requires *a lot* more energy. This information is the basis of our case against negative splitting: usually, the pacers are gone for much of the second half of the race, and the main runner already faces increased energy expenditure due to increased drag. Running that portion of the race faster would result in even higher energy consumption, which seems detrimental to performance *a priori*.

However, negative splitting means that the earlier stages of the race were run slower, which in turn results in the pacers saving more energy and staying in the race for longer (thus affording the main runner a drag benefit for extended time). These two aspects of negative splitting are counterproductive, and the optimal solution likely lies somewhere in the middle. In order to settle this question, we performed an energetic analysis of a runner and his pacer running in formation in a realistic marathon. The drag values were obtained as functions of running velocity by means of computational fluid dynamics

(CFD) and fed into a running power model. Finally, performance predictions were made for a range of running speeds, providing a definitive answer to the pacing problem.

## 2. Methods
### Obtention of drag values

For simplicity, this study considers only two running formations: the single runner case and Formation 1, where the main runner runs in a protected position 1.2 m behind the front pacer (Fig. 2). Formation 1 is chosen for its simplicity (it can be used whenever at least one pacer is available) and for its appearance in multiple races (Eliud Kipchoge employed it in both the 2018 and 2022 Berlin Marathons).

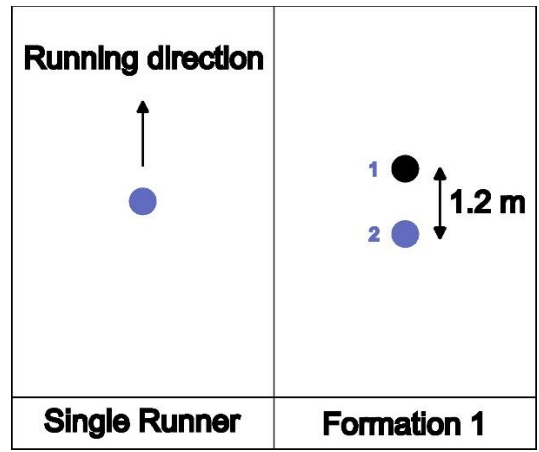

Fig. 2 Running formations considered in this study. The main runner runs in position 2 and the pacer runs in position 1.

The computational fluid dynamics (CFD) simulations were performed in previous studies [18,20,21] for both running formations using a high quality static anthropometric model, properly modified to match Kipchoge's height and running position. The results for the drag experienced by the main runner and his pacer are shown in Fig. 3.

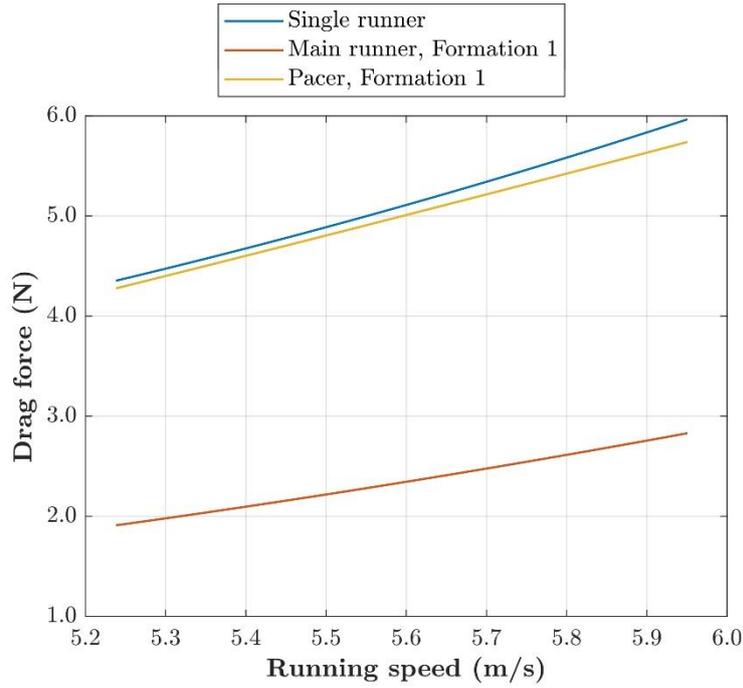

Fig. 3 Drag plots for the runners in the relevant formations. Curves shown are a second order polynomial fits of four discrete values, which are fed into the equations in section 0.

By simply running behind the pacer the main runner experiences about 53% less drag. Since the obtention of drag values is not the focus of this paper, we chose to present the results directly. For more information on the CFD procedures and analysis for these and more running formations, please refer to sources [18–21].

## Calculation of power values

Mathematical models exist for a wide variety of physical phenomena [22]. The calculation of the power output of a runner is made through one of the available empirical models. The earliest model was proposed by Fukunaga et al [23]. In this study, the more complete model, proposed by Cavagna and Kaneko [24], is used. The equation for specific running power per unit mass is

$$P_{R,0} = 9.42 + 4.73 \cdot u + 0.266 \cdot u^{1.993}$$

In this model, the velocity $u$ is in $km/h$ and $P_{R,0}$ is calculated in calories per minute per kg. To facilitate the analysis, changes were made so that the velocity is in $m/s$ and $P_{R,0}$ is found in Joules per second per kg. In this study, the runner of choice is Eliud Kipchoge, who weighs 52 kg [25]. The previous equation is multiplied by his mass to obtain the total running power, in W:

$$P_R = 34.1581 + 61.7457u + 12.3889u^{1.993}$$

This model contemplates all the components of energy expenditure except for aerodynamic resistance. This portion, known as aerodynamic power, is calculated as

$$P_A = F_D \cdot u$$

Where $F_D$ is the drag force, in $N$, and $u$ is the running velocity, in $m/s$; the aerodynamic power is obtained in $W$. The drag force, $F_d$, is obtained from the CFD simulations (section 2). The total running power, $P_{tot}$, is the sum of both components:

$$P_{tot} = P_R + P_A$$

## 3. Analytical procedure

The energetic analysis of a runner during a marathon starts with a detailed evaluation of the race broadcast and splits to determine the average speed and formation for each portion of the race. We chose to perform our analysis for the 2022 Berlin Marathon [26]. It is worth noticing that Kipchoge started the race with three pacers and adopted different formations. However, since we are only considering two formations, our analysis will focus only on Kipchoge and on his strongest pacer, i.e., the one who stayed in the race for longer.

Since the running speed and formation are known for each portion of the race, $F_D$ can be determined (Fig. 3), and the total running power can be calculated. The results are then multiplied by the time and tabulated to determine the total energy expenditure of each runner during the race. For the full results, please refer to the supplementary materials. For that specific race, Eliud Kipchoge spent $6.02 \cdot 10^6 J$ of energy, and his best pacer, Jacob Kiplimo, spent $3.6 \cdot 10^6 J$ (about 60% of Kipchoge's energy expenditure).

The next step is to divide the race into two hypothetical phases: in the first phase, Kipchoge runs behind his pacer with velocity $v_i$. This is referred to as the "protected" phase. When the pacer drops out, Kipchoge finishes the race by himself with velocity $v_f$, in the "unprotected" portion of the race. Our goal is to figure out the best way to choose velocities $v_i$ and $v_f$ while preserving the energy for both Kipchoge and his pacer, so that the race can be completed in the minimum possible time.

For a more in-depth analysis, we consider different energy values for the pacer by freely choosing a factor $k = E_{pacer}/E_{Kipchoge}$. For instance, the real race corresponds to the case $k = 0.6$. We consider $k$ values from 0.1 to 0.9, with increments of 0.1. The energetic analysis was performed using MATLAB (The script is provided in the supplementary materials). The procedure is outlined below:

For each $k$ value, from 0.1 to 0.9:

- Choose factor $k$ and calculate the energy of the pacer as a fraction of Kipchoge's energy;

For 50 velocity values, dividing the interval from 5.66 $m/s$ to 6.22 $m/s$ into equal parts:

- Choose protected velocity $v_i$. Determine how much distance is covered and time is elapsed until the pacer extinguishes his energy by calculating the pacer's total running power, $P_{tot}$.

- Calculate Kipchoge's total running power and determine how much energy is spent during the protected phase, as well as how much energy remains for the unprotected phase.
- Declare the unprotected velocity $v_f$ as a variable and determine how much energy (and time) is spent to cover the remaining distance as a function of $v_f$, $E(v_f)$.
- Apply conservation of energy by forcing equality between the remaining energy available and $E(v_f)$. Numerically solve the equation to find $v_f$ and calculate the overall race time.
- Plot the race times against $v_i$ for all possible cases.

## 4. Results and discussion

The race time results for varying protected velocity ($v_i$) and $k$ values are shown in Fig. 4. For $k = 0.1$, the choice of $v_i$ influences the total race time very little. This is an expected result since the pacer does not have enough energy to stay in the race long enough to benefit the main runner significantly. As the value of $k$ increases, the pacer has enough energy to affect the pacing strategy. For the scenario that occurred in the actual race ($k = 0.6$), the curve has a minimum of 7180 seconds (1:59:40) of race time at around $v_i = 5.94 \ m/s$. This combination of drafting and pacing strategy could have allowed Kipchoge to break the 2-hour barrier. The higher values of $k$ are unrealistic for real race situations, but are included here for illustration purposes.

The shape of the curves allows us to illustrate the counterproductive aspects of pacing cited in section 1. On one hand, if the protected velocity is too low, the pacer stays in the race longer (allowing the main runner to preserve more energy), however running the final portion faster increases the energy expenditure a lot. Since the total energy expenditure is fixed, this strategy results in higher race times. On the other hand, if the protected velocity is too high, the pacer drops early, affording the main runner less aerodynamic protection. By experiencing higher drag, the main runner has to slow down to balance his energy expenditure, again resulting in higher race times.

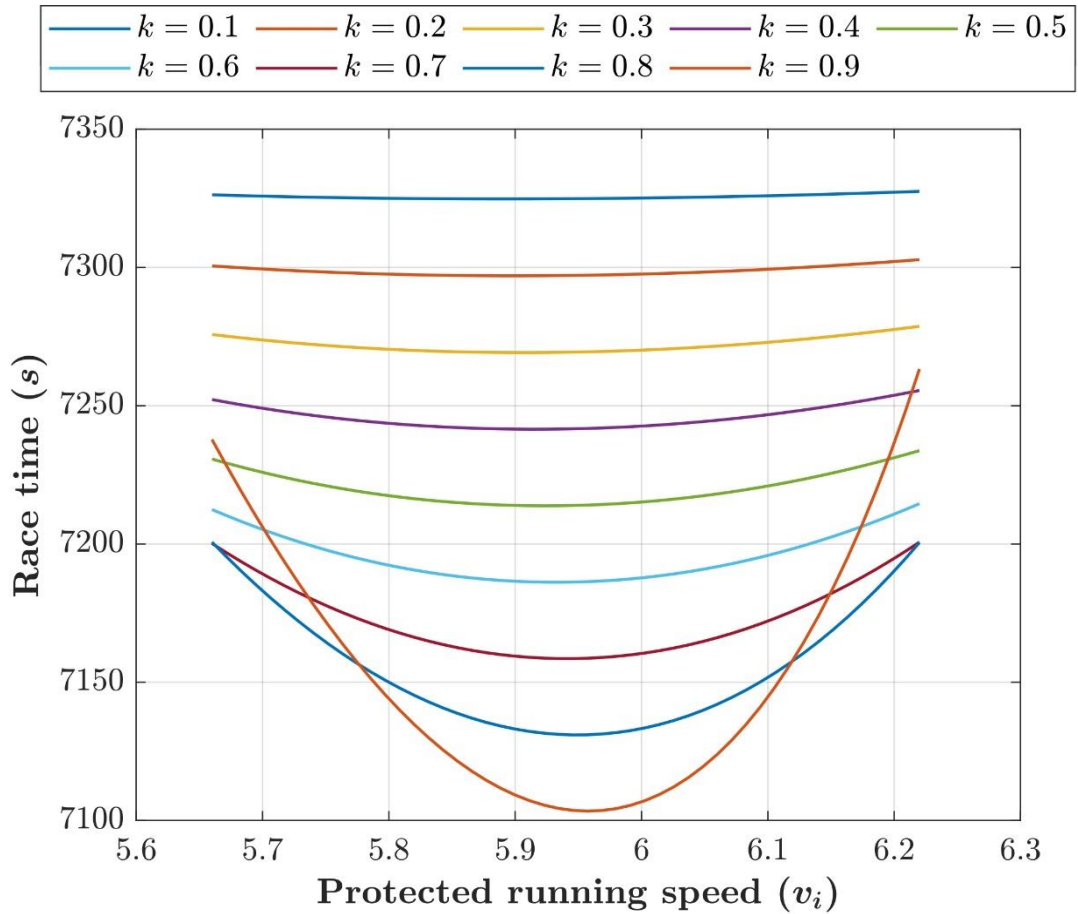

Fig. 4 Total race time as a function of protected running speed, $v_i$, for multiple $k$ values.

The optimal result is indeed found somewhere in the middle. Regardless of the value of $k$, the minimum race time always happens when the protected phase is run at a speed between 5.9 and 6 $m/s$. For instance, an ideal 2-hour marathon is run at a steady speed of 5.86 $m/s$. It is clear that the first portion of the race must be run faster in order to achieve minimum time, highlighting the advantage of positive splitting whenever pacers are available.

The relationship between protected and unprotected running speed for the optimal races is shown in Fig. 5. Indeed, for the family of races with shorter time, the protected phase is run faster. Additionally, the extent to which a positive pacing strategy should be employed (i.e., *how much faster should the protected phase be run*) appears to increase with $k$. In other words, the more energy the pacer is able to develop, the faster the protected speed should be.

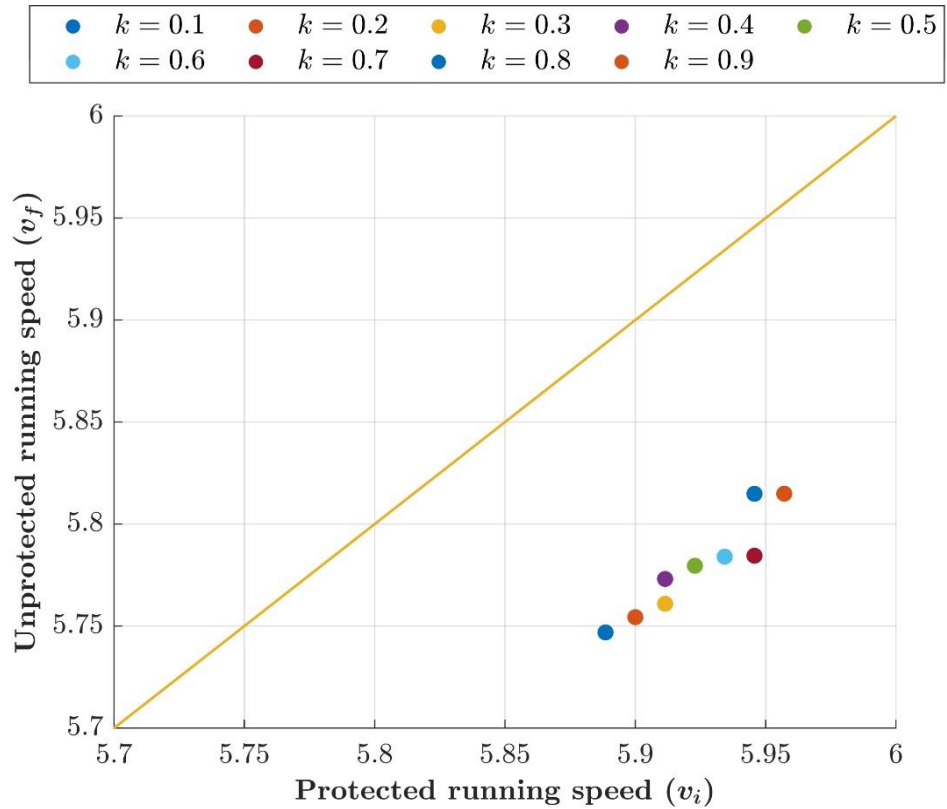

Fig. 5 Protected and unprotected running speeds for optimal races. The red line represents even pacing. The area to its right represents positive splitting.

Nevertheless, for all cases, the amount of positive splitting is very slight, with the protected velocity $v_i$ always between 2.4 and 2.6% faster than the unprotected velocity $v_f$ for the optimal races. Our results show that, when pacers are part of the race effort and properly used to reduce the drag experienced by the main runner, the best pacing strategy is indeed positive splitting.

## 5. Limitations

The present study explored a simple combination of drafting/pacing effort to achieve optimal marathon performance. In real race situations, runners often have more than one available pacer, and there are more aerodynamically efficient formations can be used. Each race situation has its own particularities, and the optimal drafting/pacing strategy will vary. Also, it should be noted that it is very difficult for any runner to maintain his or her pace perfectly on target, so our results should be seen as a guideline for the target pacing.

## 6. Conclusion

Our results provide evidence that the presence of pacers indeed affects the optimal pacing strategy for elite marathon runners. Contrary to popular belief that a negative splitting strategy is better, it is evident that, in this case, a strategy adopting slightly positive pacing is optimal.

**Statements and disclosures**

**Funding information**: no funding was used in this research project.

**Author contributions**: both authors have contributed in all phases of the project, as well as in the preparation of the manuscript.

**Competing interests**: the authors declare no conflicts of interest.

**Data sharing**: all the relevant data is included in the manuscript and in the supplementary materials.